\documentclass[aps,prl,twocolumn,groupedaddress,showpacs]{revtex4}
\usepackage{bm}
\usepackage{graphicx}
\usepackage{dcolumn}
\usepackage{color}

\begin{document}

\title{Relationship between Fermi-Surface Warping and Out-of-Plane Spin Polarization in Topological Insulators: 
a View from Spin-Resolved ARPES}
\author{M. Nomura,$^1$ S. Souma,$^2$ A. Takayama,$^2$ T. Sato,$^1$ T. Takahashi,$^{1,2}$\\
 K. Eto,$^3$ Kouji Segawa,$^3$ and Yoichi Ando$^3$}
\affiliation{$^1$Department of Physics, Tohoku University, Sendai 980-8578, Japan}
\affiliation{$^2$WPI Research Center, Advanced Institute for Materials Research, 
Tohoku University, Sendai 980-8577, Japan}
\affiliation{$^3$Institute of Scientific and Industrial Research, Osaka University, Ibaraki, Osaka 567-0047, Japan}

\date{\today}

\begin{abstract}
We have performed spin- and angle-resolved photoemission spectroscopy of the topological insulator Pb(Bi,Sb)$_2$Te$_4$ (Pb124) and observed significant out-of-plane spin polarization on the hexagonally warped Dirac-cone surface state. To put this into context, we carried out quantitative analysis of the warping strengths for various topological insulators (Pb124, Bi$_2$Te$_3$, Bi$_2$Se$_3$, and TlBiSe$_2$) and elucidated that the out-of-plane spin polarization $P_z$ is systematically correlated with the warping strength. However, the magnitude of $P_z$ is found to be only half of that expected from the $k{\cdot}p$ theory when the warping is strong, which points to the possible role of many-body effects. Besides confirming a universal relationship between the spin polarization and the surface state structure, our data provide an empirical guiding principle for tuning the spin polarization in topological insulators.
\end{abstract}

\maketitle

Three-dimensional topological insulators (TIs) realize a novel quantum state of matter where a distinct topology of bulk wave functions produces a gapless surface state (SS) which disperses across the bulk insulating gap \cite{HasanRMP, ZhangRMP, AndoRev}. The topological SS is characterized by a Dirac-cone band dispersion with an in-plane spin helical texture arising from the strong spin-orbit coupling (SOC). This peculiar spin texture plays an essential role in realizing various quantum phenomena associated with non-trivial topology \cite{QiPRB, QiScience, FuKaneMajorana}, and also in manipulating spin polarization and spin current in TIs \cite{Louie}.

It is known that the spin texture of the Dirac-cone SS is closely related to the shape of the Fermi surface (FS) \cite{FuPRL, Model}, as proposed by the $k{\cdot}p$ perturbation theory for TIs with $C_{3v}$ symmetry \cite{FuPRL}. Theoretically, the $k$ cubic term in the surface SOC causes a hexagonal warping of the FS as well as a finite out-of-plane spin polarization ($P_z$) with $C_3$ symmetry. Such a deviation from an ideal isotropic Dirac cone has been thought to cause a variety of intriguing physical properties. For instance, the FS warping is responsible for the strong quasiparticle interference as seen by scanning tunneling microscopy \cite{XueSTM, KapitulnikSTM}. Also, the warping could trigger the spin density wave and create an energy gap at the Dirac point under transverse magnetic field \cite{FuPRL}.

The out-of-plane spin polarization has been observed by spin- and angle-resolved photoemission spectroscopy (ARPES) for Bi$_2$Te$_3$ \cite{SomaBT, HasanBT} in which the pronounced FS warping leads to the finite $P_z$ component along the $\bar{\Gamma}$-$\bar{K}$ direction in the surface Brillouin zone (BZ). While previous spin-resolved ARPES experiments suggested a correspondence between the FS warping and the out-of-plane spin polarization, no systematic investigation on the relationship has been performed. This is largely due to the small magnitude of the $P_z$ value compared to the in-plane counterpart and the limitation of materials suitable for observing the out-of-plane spin polarization. In addition, the spin-polarization values reported from spin-resolved ARPES have not been very consistent between experiments even for the same materials \cite{SomaBT, HasanBT, HsiehNatBT, VallaBiSe, JozwiakPRB}. Thus, systematic spin-resolved ARPES experiments for various TIs with an identical experimental condition are highly desirable for elucidating the universal relationship between the spin polarization and the FS warping in TIs.

 In this Letter, we report the spin polarization and the energy band dispersion of the Dirac-cone SS for various TIs. We observed a pronounced out-of-plane spin polarization in PbBi$_2$Te$_4$ and Bi$_2$Te$_3$, in sharp contrast to Bi$_2$Se$_3$ and TlBiSe$_2$. We extracted the FS-warping strength by numerical fittings to the experimental band dispersions and found a universal relationship between the out-of-plane spin component and the FS warping, which is essential for understanding the dynamics of Dirac fermions in TIs.

  High-quality single crystals of Pb(Bi,Sb)$_2$Te$_4$ (called Pb124 here), Bi$_2$Se$_3$, Bi$_2$Te$_3$, and TlBiSe$_2$ (Tl112)  were grown by a modified Bridgman method from high-purity elements sealed in an evacuated quartz-glass tube \cite{SomaPb, SomaBT, SatoPRL}. Spin-resolved and regular ARPES measurements were performed using an ultrahigh-resolution spin-resolved ARPES spectrometer based on the MBS-A1 analyzer at Tohoku University \cite{SoumaRSIspin}. We used one of the Xe I resonance lines ($h\nu$ = 8.437 eV)  to excite photoelectrons. ARPES measurements were also performed at the beam line BL28A at Photon Factory (KEK), where we used circularly polarized light of 36-80 eV. All the samples were cleaved $in$-$situ$ along the (111) crystal plane in an ultrahigh vacuum of 5-9$\times10^{-11}$ Torr. The energy resolutions during the spin-resolved and regular ARPES measurements were 20-40 and 6-10 meV, respectively. We used the Sherman function value of 0.07 for the spin-resolved ARPES measurements.  The Fermi level ($E_{\rm F}$) of the samples was referred to that of a gold film evaporated onto the sample holder.

\begin{figure}
\includegraphics[width=3in]{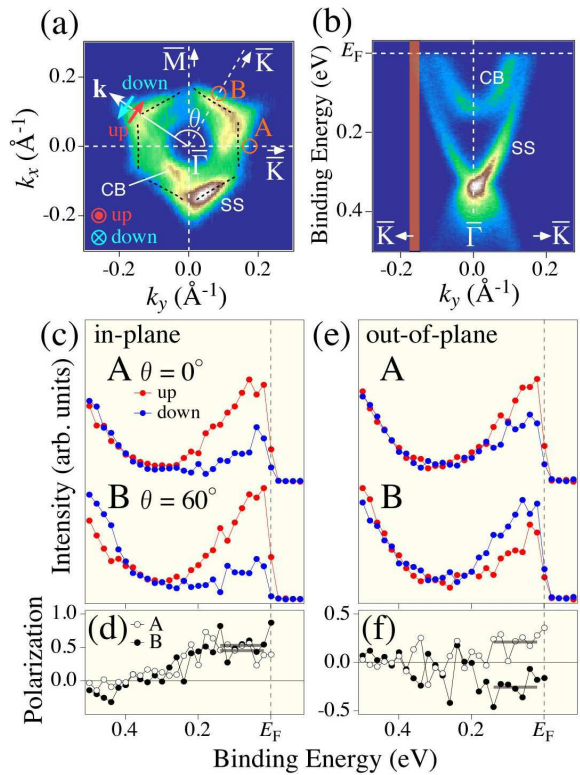}
\vspace{-0.4cm}
\caption{(color online): (a) ARPES intensity at $E_{\rm F}$ of Pb124 around the $\bar{\Gamma}$ point plotted as a function of two-dimensional wave vector $\bf{k}$. The definition of the FS angle $\theta$ and up/down-spin vectors for the in-plane and out-of-plane components is also indicated. Black dashed curves are guides to the eye to trace the surface state (SS). For the present spin-resolved ARPES measurements, we chose Pb(Bi$_{0.6}$Sb$_{0.4}$)$_2$Te$_4$, which is less electron doped compared to PbBi$_2$Te$_4$ in a rigid band manner \cite{SomaPb}, to reduce the contamination from the bulk conduction band (CB). (b) Near-$E_{\rm F}$ ARPES intensity as a function of wave vector along the $\bar{\Gamma}$-$\bar{K}$ line and the binding energy. (c) Spin-resolved EDCs for the in-plane component at $\bf{k}$ points A ($\theta=0^{\circ}$) and B ($\theta=60^{\circ}$) indicated by open circles in (a). The $k$ region for the spin-resolved measurements is indicated by thick vertical line in (b). (d) In-plane spin polarization at points A and B. (e), (f) Similar plots for the out-of-plane spin component. Thick horizontal lines in (d) and (f) represent the averaged spin polarization near $E_{\rm F}$.}
\end{figure}

First we demonstrate the band structure and spin polarization for Pb124 in which the pronounced hexagonal FS warping was reported in previous ARPES studies \cite{SomaPb, HiroPb}. Figure 1(a) displays the ARPES-intensity map at $E_{\rm F}$ around the $\bar{\Gamma}$ point as a function of the in-plane wave vector $\bf{k}$. We immediately notice a triangular intensity pattern centered at the $\bar{\Gamma}$ point, which originates from both the three-fold symmetric bulk conduction band (CB) and the hexagonally-warped gapless SS \cite{SomaPb}.  As shown in the representative band dispersion along the $\bar{\Gamma}$-$\bar{K}$ cut in Fig. 1(b), the SS can be distinguished from the bulk CB, enabling us to determine accurately the Fermi vectors ($k_{\rm F}$) of the SS [black dashed curves in Fig. 1(a)].

We have measured spin-resolved ARPES spectra on the $\bar{\Gamma}$-$\bar{K}$ line where the largest out-of-plane spin component is expected \cite{SomaBT, FuPRL, HasanBT}. We also chose the particular $\bf{k}$ point in which the peak structure at $E_{\rm F}$ is dominated by the SS [thick vertical line in Fig. 1(b)]. Figure 1(c) shows the spin-resolved energy dispersion curves (EDCs) for the in-plane component at points A and B in the BZ [Fig. 1(a)]. We define the in-plane spin polarization vector to be perpendicular to the measured $\bf{k}$, and the ``up-spin" points to the clockwise direction in Fig. 1(a). Obviously, the up-spin intensity is stronger than the down-spin counterpart at both points A and B, indicating the overall spin-helical texture [the spin polarization averaged around $E_{\rm F}$ is $\sim$ 0.5; see Fig. 1(d)]. As shown in Fig. 1(e), we also identify a large difference between the up- and down-spin EDCs for the out-of-plane component, whereupon the relative intensity for the up and down spins are reversed between the two points (A and B) indicating that the overall spin texture has the $C_3$ symmetry \cite{SomaBT}, consistent with the theoretical prediction \cite{FuPRL}. We also note that the averaged magnitude of the out-of-plane spin polarization, $|P_z|$, is $\sim$ 0.2 [see Fig. 1(f)], comparable to that of Bi$_2$Te$_3$ \cite{SomaBT, HasanBT}.

\begin{figure}
\includegraphics[width=3.4in]{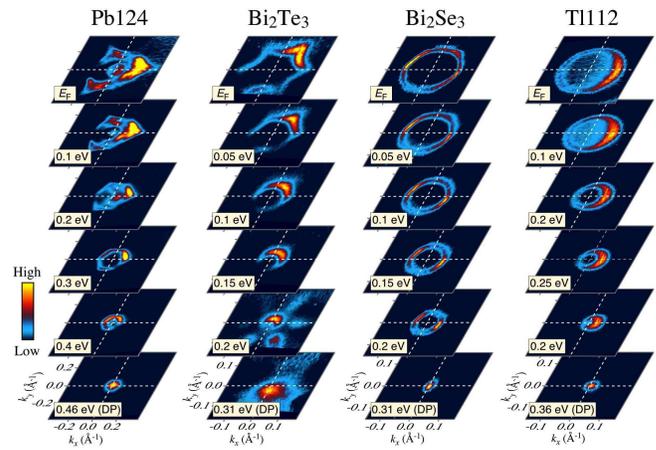}
\vspace{-0.4cm}
\caption{(color online): ARPES intensity mappings for Pb124 (PbBi$_2$Te$_4$), Bi$_2$Te$_3$, Bi$_2$Se$_3$, and Tl112 as a function of 2D wave vector at various binding energies. DP denotes the Dirac point.}
\end{figure}

To obtain further insight into the relationship between the out-of-plane spin polarization and the FS warping, it would be useful to determine the warping strength for various TIs through accurate band structure mappings. Figure 2 shows the two-dimensional (2D) energy contour plots of the ARPES intensity around the $\bar{\Gamma}$ point for Pb124, Bi$_2$Te$_3$, Bi$_2$Se$_3$, and Tl112, all of which exhibit a single Dirac-cone SS centered at the $\bar{\Gamma}$ point \cite{HasanBi2Se3, ShenBT, HsiehNatBT, SatoPRL, HiroTl, ChenTl, SomaPb, HiroPb, EremeevPb}. Owing to the inherently electron-doped nature of these compounds, the energy position of the Dirac point (Dirac energy) is located well below $E_{\rm F}$ (0.31 - 0.46 eV), which can be recognized from the gradual shrinkage of the intensity as the energy is lowered from $E_{\rm F}$. A closer look at Fig. 2 further reveals that the shape of the intensity distribution is very much material-dependent; the intensity pattern of Bi$_2$Te$_3$ is snowflake-like at the binding energy away from the Dirac point \cite{ShenBT, SomaBT}, while that for Bi$_2$Se$_3$ and Tl112 is ring-like. For Pb124, a hexagonal pattern is clearly seen at 0.3 eV, above which the intensity is strongly modulated by the bulk CB [see also Figs. 1(a) and (b)].

\begin{figure}
\includegraphics[width=3.4in]{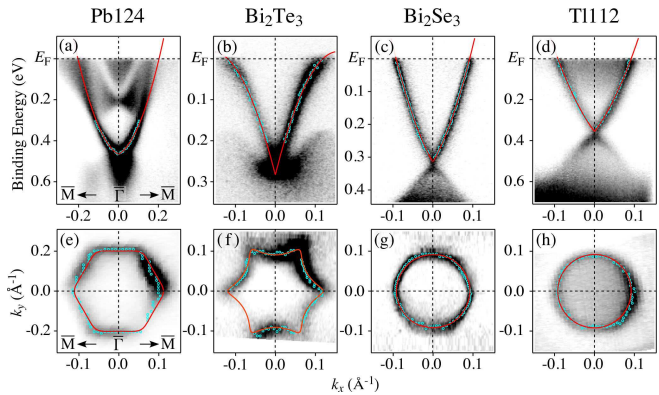}
\vspace{-0.4cm}
\caption{(color online): (a)-(d) Comparison of the band dispersions near $E_{\rm F}$ along the $\bar{\Gamma}$-$\bar{M}$ cut for Pb124 (PbBi$_2$Te$_4$), Bi$_2$Te$_3$, Bi$_2$Se$_3$, and Tl112. (e)-(h) Energy contour plots of the 2D ARPES intensity; the binding energy of the contour plot for Pb124 was chosen to be 0.3 eV to avoid contamination from the bulk CB, while that for other samples was fixed at $E_{\rm F}$. Band dispersion and the exact location of the equi-energy contour (light-blue dots) are determined from the peak position in the momentum distribution curves. Red solid curves are the results of numerical fittings to the obtained band dispersions and the equi-energy contours with Eq. (1) \cite{FuPRL}.
}
\end{figure}

To evaluate quantitatively the warping strength, we numerically fit the experimental band structure for the upper Dirac cone with the theoretical $E$-$\bf{k}$ relationship deduced from the $k{\cdot}p$ theory \cite{FuPRL, Model},

\begin{eqnarray}
E(k,\theta)=E_{\rm DP} + k^2/2m^{*} + \sqrt{v^2k^2 + {\lambda}^2k^6{\rm cos}^2(3\theta)},
\label{eq:one}
\end{eqnarray}
where $k$ is the magnitude of in-plane wave vector $\bf{k}$, and $\theta$ is the angle with respect to the $\bar{\Gamma}$-$\bar{K}$ line, as shown in Fig. 1(a). $E_{\rm DP}$, $m^{*}$, and $v$ are the Dirac energy, the ``effective mass" term to represent the curvature, and the Dirac velocity, respectively. $\lambda$ is the coefficient of the $k^3$ SOC term which represents the warping strength. First, we fit the Dirac-cone energy dispersion along the $\bar{\Gamma}$-$\bar{M}$ line ($\theta=90^{\circ}$), in which the contribution from the warping term is zero, to determine the parameters $E_{\rm DP}$, $m^{*}$, and $v$. Then, we estimated the warping strength $\lambda$, by fitting the $k_F$ points in 2D $\bf{k}$ space (for Pb124, the $k$ point is at 0.3 eV) by using the $E_{\rm DP}$, $m^{*}$, and $v$ values obtained as above. As shown in Fig. 3, the fitted band dispersions and energy contour plots (red curves) show good agreements with the experimental data, indicating the basic validity of the  $k{\cdot}p$ theory in describing the band structure of the Dirac-cone SS. It is also noted that the energy dispersion in Figs. 3(a)-(d) is not linear at $k$ away from the $\bar{\Gamma}$ point even for the weakly warped TIs Bi$_2$Se$_3$ and Tl112, due to the finite contribution from the $m^{*}$ term \cite{Taskin}.

We show the obtained fitting parameters for various TIs in Table I, where we find that the warping strength $\lambda$ significantly varies in different materials. Bi$_2$Te$_3$ shows the largest value of 210 eV$\rm{\AA^3}$ while Tl112 has a negligible value (0 eV$\rm{\AA^3}$), reflecting the shapes of the FS (snowflake or circle, respectively). One may wonder why similar $\lambda$ values were obtained for Pb124 and Bi$_2$Se$_3$ despite the marked difference in the FS shape (hexagonal or circular). This is due to the smaller Dirac velocity $v$ in Pb124 which enhances the warping effect, because the warping strength is characterized by the ratio between the ${\lambda}^2k^6{\rm cos}^2(3\theta)$ term and the $v^2k^2$ term.  Here, we introduce another parameter $a=\sqrt{\lambda/v}$ more suitable for quantitative discussions of the intrinsic warping strength \cite{FuPRL}. As seen in Table I, it is clear that the new parameter $a$ exhibits a more direct correspondence to the observed FS shape.

\begin{table}
\caption{\label{tab:table1}${m^*/m}, v, \lambda$, and $a=\sqrt{\lambda/v}$, obtained from the numerical fittings to the ARPES data for Pb124, Bi$_2$Te$_3$, Bi$_2$Se$_3$, and Tl112.}
\begin{ruledtabular}
\begin{tabular}{lcccc}
 &Pb124 & Bi$_2$Te$_3$ & Bi$_2$Se$_3$ &Tl112\\
\hline
${m^*/m}$& 0.4 & -0.3 & 0.3 &0.3\\
$v$ [eV$\rm{\AA}$]& 0.5 & 3.8 & 2.1 &2.5\\
$\lambda$ [eV$\rm{\AA^3}$]& 55 & 210 & 50 &0\\
$a=\sqrt{\lambda/v}$ [$\rm{\AA}$]& 10.6 & 7.5 & 4.8 &0.0\\
\end{tabular}
\end{ruledtabular}
\end{table}

   Now that the warping strength is quantitatively established, the next issue to be elucidated is the material dependence of the out-of-plane spin polarization.  In addition to Pb124, we have performed spin-resolved ARPES measurements for Bi$_2$Se$_3$, Bi$_2$Te$_3$, and Tl112 around the $k_F$ point along the $\bar{\Gamma}$-$\bar{K}$ line ($\theta=0^{\circ}$), with the same experimental condition as in Pb124 (Fig. 1). We show in Fig. 4(a) the obtained $P_z$ values at $\theta=0^{\circ}$ for various TIs averaged over a narrow energy window near $E_{\rm F}$, together with the data from previous spin-resolved ARPES measurements for Bi$_2$Te$_3$ \cite{HasanBT} and PbBi$_4$Te$_7$ (Pb147) \cite{EremeevPb}. Since the out-of-plane spin polarization depends on the distance between the measured $\bf{k}$ and the $\bar{\Gamma}$ point \cite{SomaBT, FuPRL}, we plot the $P_z$ value as a function of $ka$ instead of $a$ to take this factor into account. Note that $a$ is not the lattice constant but is $\sqrt{\lambda/v}$ to measure the warping strength. As shown in Fig. 4(a), the $P_z$ value gradually increases with increasing $ka$ of up to $\sim 1.0$, and then tends to saturate at larger $ka$. To compare this behavior with theory, we have calculated the theoretical $P_z$ value using the $k{\cdot}p$ theory \cite{FuPRL}, 

\begin{eqnarray}
P_z(k,\theta)={\rm cos}(3\theta)/{\sqrt{{\rm cos}^2(3\theta) + 1/(ka)^4}}
\label{eq:two}
\end{eqnarray}
and plotted the obtained value for $\theta=0^{\circ}$ (the $\bar{\Gamma}$-$\bar{K}$ line) in Fig. 4(a). One finds that the $P_z$ value from the $k{\cdot}p$ theory is much larger than that of the experiment. This may partly be due to the contribution from the non-polarized bulk CB \cite{MeierNJP, VallaBiSe} and/or the possible spin-orbit entanglement \cite{Louie}. These extrinsic effects can be removed by dividing the $P_z$ value by the total spin polarization $P$, because (i) the bulk CB affects equally both the $P_z$ and $P$ values and (ii) the spin-orbit entanglement simply reduces the total magnitude of the spin polarization \cite{Model} and does not affect the direction of the spin vector. Figure 4(b) compares the theoretical and experimental $P_z/P$ value (which signifies the intrinsic out-of-plane spin polarization) plotted as a function of $ka$. We still find the theoretical $P_z/P$ value to be much larger than the experimental value in the large $ka$ regime ($ka > 0.5$), even after the extrinsic effects are eliminated. There is roughly a factor of two difference between theory and experiment at large $ka$.

\begin{figure}[t]
\includegraphics[width=3in]{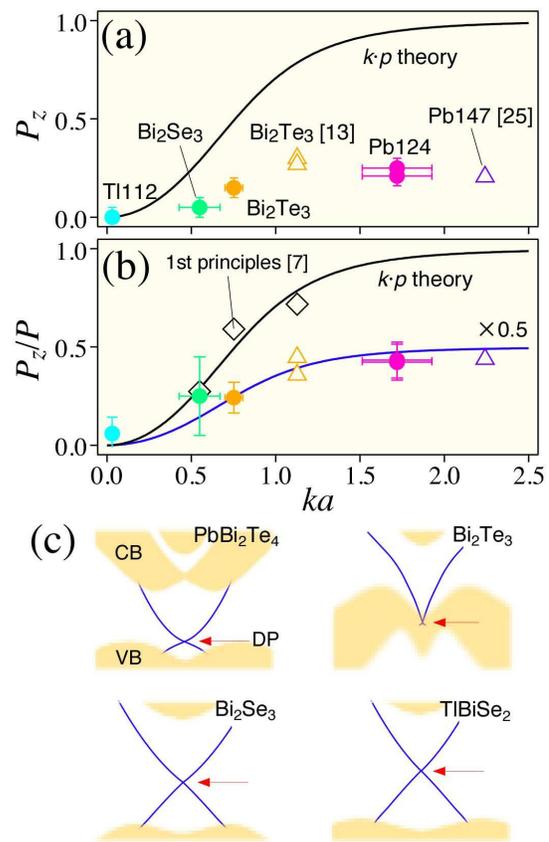}
\caption{(Color online): (a) Experimentally obtained $P_z$ values (for Pb124, Bi$_2$Se$_3$, Bi$_2$Te$_3$, and Tl112) as a function $ka$, together with the data from previous spin-resolved ARPES for Bi$_2$Te$_3$ \cite{HasanBT} and PbBi$_4$Te$_7$ (Pb147) \cite{EremeevPb}. Solid curve is the theoretical $P_z$ value for $\theta=0^{\circ}$ (the $\bar{\Gamma}$-$\bar{K}$  line) calculated from Eq. (2) \cite{FuPRL}. (b) Similar plot with the vertical axis changed to $P_z/P$, where $P$ is the total spin polarization expressed as $P = \sqrt{P_x^2+P_y^2+P_z^2}$ with $P_x$ and $P_y$ the in-plane spin polarizations parallel and perpendicular to the $\bar{\Gamma}$-$\bar{K}$ line, respectively (the $P_x$ value is zero along the $\bar{\Gamma}$-$\bar{K}$ line due to the mirror symmetry of the crystal). The $P_z/P$ values from the first-principle calculations for Bi$_2$Se$_3$ and Bi$_2$Te$_3$ \cite{Louie} are also plotted with open diamonds. (c) Comparison of the schematic energy band structure near $E_{\rm F}$ for various TIs. Blue curves represent the Dirac-cone SS while the yellow shades denote the continuum of the bulk CB and VB. Red arrow indicates the energy position of the Dirac point (DP).
}
\end{figure}

We remark here that it is not clear in our data whether the factor of two difference also exists at small $ka$ ($\lesssim$ 0.5), due to the relatively large experimental uncertainty in the Bi$_2$Se$_3$ data where $P_z$ is very small. In this regard, a recent circular-dichroism ARPES measurement for Bi$_2$Se$_3$ \cite{CDARPES} found that the out-of-plane spin polarization in this material is well described by the $k{\cdot}p$ theory, which suggests that the deviation of the experimental data from the theoretical prediction regarding the magnitude of $P_z$ appears only at large $ka$.  This is probably reasonable, because the $k{\cdot}p$ theory is an approximation applicable to a narrow $k$ region near the BZ center (note that inclusion of a higher order warping term \cite{Bansil} may extend its applicability toward higher $k$ region). 

In the larger $ka$ regime, it is useful to note that the $P_z/P$ values obtained from first-principle band calculations for Bi$_2$Te$_3$ \cite{Louie} [see open diamonds in Fig. 4(b) at $ka$ = 0.75 and 1.1] agree with the $k{\cdot}p$ theory but not with experiments. This fact points to the possibility that the reduction in the actual out-of-plane spin polarization at large $ka$ may originate from the physics beyond the simple one-particle approximation, such as many-body effects and/or the instability toward spin density wave \cite{FuPRL}.

Having established the universal relationship between the intrinsic out-of-plane spin polarization and the warping parameter, an interesting question is whether it is possible to tune the warping parameter $\lambda$, which may open a way for engineering the spin polarization in TIs. Theoretically, the warping term for the SS stems from the cubic Dresselhaus SOC term of the bulk states  \cite{FuPRL,Model}; the coefficients of these terms are related to each other through the equation $\lambda = \alpha{R}$, where $\alpha$ is a scaling factor and $R$ is the coefficient of the Dresselhaus SOC term. The theoretical $R$ value for Bi$_2$Se$_3$ and Bi$_2$Te$_3$ is 50 and 45 $\rm{\AA^3}$, respectively \cite{Model}, which correspond to the $\alpha$ value of 1 and 4.6 (estimated from the $\lambda$ values in Table I). This demonstrates that the cubic Dresselhaus SOC term is not equal to the surface warping parameter itself because of the material-dependent scaling factor $\alpha$. While the origin of $\alpha$ has not been explicitly discussed so far, we speculate that it is related to the energy location of the Dirac point relative to the bulk state. As shown in Fig. 4(c) which depicts the band picture for various TIs, there is a general trend that the Dirac point in Te-based materials is buried in the bulk valence band while that in Se-based materials is isolated from the bulk state. Hence, empirically it seems that the scaling factor $\alpha$ becomes larger when the Dirac point is closer to the bulk valence band. If it is actually the case, it may be possible to tune the FS warping and the resulting spin polarization by controlling the energy location of the Dirac point through Se/Te alloying.

 In summary, we have reported spin-resolved ARPES experiments on various TIs to experimentally establish the relationship between the out-of-plane spin polarization and the FS warping. We confirmed a universal trend that stronger FS warping leads to a larger out-of-plane spin polarization, although the intrinsic out-of-plane spin polarization is reduced by a factor of 2 compared to that expected from the $k{\cdot}p$ theory when the warping is strong, which points to the possible role of many-body effects in the topological surface state.

\begin{acknowledgments}
We thank L. Fu for stimulating discussions. We also thank K. Nakayama, Y. Tanaka, T. Shoman for their help in the ARPES measurements and M. Kriener for his help in crystal growths. This work was supported by JSPS (KAKENHI 23224010, 24654096, and 25220708), MEXT of Japan (Innovative Area gTopological Quantum Phenomenah), AFOSR (AOARD 124038), Mitsubishi foundation, and KEK-PF (Proposal number: 2012S2-001). 
\end{acknowledgments}


\begin{references}
\bibitem{HasanRMP}
M. Z. Hasan and  C. L. Kane, Rev. Mod. Phys. {\bf 82}, 3045 (2010).
\bibitem{ZhangRMP}
X.-L. Qi and S.-C. Zhang, Rev. Mod. Phys. {\bf 83}, 1057 (2011).
\bibitem{AndoRev}
Y. Ando, arXiv:1304.5693.
\bibitem{QiPRB}
X.-L. Qi, T. L. Hughes, and S.-C. Zhang, Phys. Rev. B {\bf 78}, 195424 (2008).
\bibitem{QiScience}
X.-L. Qi {\it et al.}, Science {\bf 323}, 1184 (2009).
\bibitem{FuKaneMajorana}
L. Fu and C. L. Kane, Phys. Rev. Lett. {\bf 100}, 096407 (2008).
\bibitem{Louie}
O. V. Yazyev, J. E. Moore, and S. G. Louie, Phys. Rev. Lett. {\bf 105}, 266806 (2010).
\bibitem{FuPRL}
L. Fu, Phys. Rev. Lett. {\bf103}, 266801 (2009).
\bibitem{Model}
C. X. Liu {\it et al.}, Phys. Rev. B {\bf82}, 045122 (2010).
\bibitem{XueSTM}
T. Zhang {\it et al.}, Phys. Rev. Lett. {\bf103}, 266803 (2009). 
\bibitem{KapitulnikSTM}
Z. Alpichshev {\it et al.}, Phys. Rev. Lett. {\bf104}, 016401 (2010).
\bibitem{SomaBT}
S. Souma  {\it et al.}, Phys. Rev. Lett. {\bf106}, 216803 (2011).
\bibitem{HasanBT}
S.-Y. Xu {\it et al.},  arXiv:1101.3985.
\bibitem{HsiehNatBT}
D. Hsieh {\it et al.}, Nature {\bf460}, 1101 (2009).
\bibitem{VallaBiSe}
Z.-H. Pan {\it et al.}, Phys. Rev. Lett. {\bf106}, 257004 (2011).
\bibitem{JozwiakPRB}
C. Jozwiak {\it et al.}, Phys. Rev. B {\bf84}, 165113 (2011).
\bibitem{SomaPb}
S. Souma {\it et al.}, Phys. Rev. Lett. {\bf108}, 116801 (2012).
\bibitem{SatoPRL}
T. Sato  {\it et al.}, Phys. Rev. Lett. {\bf 105}, 136802 (2010).
\bibitem{SoumaRSIspin}
S. Souma {\it et al.}, Rev. Sci. Instrum. {\bf81}, 095101 (2010).
\bibitem{HiroPb}
K. Kuroda {\it et al.}, Phys. Rev. Lett. {\bf108}, 206803 (2012).
\bibitem{HasanBi2Se3}
Y. Xia {\it et al.}, Nature Phys. {\bf5}, 398 (2009).
\bibitem{ShenBT}
Y. L. Chen {\it et al.}, Science {\bf 325}, 178 (2009).
\bibitem{HiroTl}
K. Kuroda {\it et al.}, Phys. Rev. Lett. {\bf 105}, 146801 (2010).
\bibitem{ChenTl}
Y. L. Chen {\it et al.}, Phys. Rev. Lett. {\bf105}, 266401 (2010).
\bibitem{EremeevPb}
S. V. Eremeev {\it et al.}, Nature Commun. {\bf3}, 635 (2012).
\bibitem{Taskin}
A. A. Taskin and Y. Ando, Phys. Rev. B {\bf84}, 035301 (2011).
\bibitem{MeierNJP}
 F. Meier, J. H. Dil, and J. Osterwalder, New J. Phys. {\bf11}, 125008 (2009).
 \bibitem{CDARPES}
 Y. H. Wang {\it et al.}, Phys. Rev. Lett. {\bf107}, 207602 (2011).
\bibitem{Bansil}
S. Basak {\it et al.}, Phys. Rev. B {\bf84}, 121401(R) (2011).
\end{references}
\end{document}